\documentclass{jetpl}
\twocolumn

\lat

\newcommand\ket[1]{\left|#1\right\rangle}
\newcommand\bra[1]{\left\langle #1\right|}
\newcommand\braket[2]{\langle #1|#2\rangle}
\def\bB{{\bf B}}
\def\bn{{\bf n}}
\def\bS{\mbox{\boldmath $\hat\sigma$}}
\def\bomega{\mbox{\boldmath $\omega$}}
\def\half{{\scriptstyle\frac{1}{2}}}

\title{Geometric phase via adiabatic manipulations of the environment}
\rtitle{Geometric phase via adiabatic \ldots}
\sodtitle{Geometric phase via adiabatic manipulations of the environment}

\author{S.V.~Syzranov$^{+}$,
Yu.~Makhlin$^{*,\nabla}$\/\thanks{e-mail: sergey.syzranov@rub.de, makhlin@itp.ac.ru}}
\rauthor{S.V.~Syzranov, Yu.~Makhlin}
\sodauthor{Syzranov, Makhlin}
\address{$^+$Theoretische Physik III, Ruhr-Universit{\"a}t Bochum, 44801 Bochum, Germany\\
$^*$Landau Institute for Theoretical Physics, Kosygin str. 2, 119334, Moscow, Russia\\
$^\nabla$Moscow Institute of Physics and Technology, 141700, Dolgoprudny, Moscow region, Russia}

\abstract{We show that geometric phases may be generated in a quantum system subject to noise by adiabatic manipulations of the fluctuating fields, e.g., by variation of the system-environment coupling. For a two-state quantum system we express this phase in terms of the geometry of the path, traversed by the slowly varying direction and amplitude of the fluctuations. We discuss the origin of this phase and possibilities to separate it from the known environment-induced modification of the Berry phase.}


\begin{document}

\maketitle

{\it Introduction.}
After the discovery of the geometric phases in coherent quantum systems~\cite{Berry,ShapereWilczek} (see also~\cite{UFNBerry} on geometric phases in earlier work), it was natural to ask, whether these phases can be observed in quantum systems coupled to an environment. In particular, since the environment typically has a continuous spectrum, the gap in the spectrum of the system+environment vanishes, which blocks manipulations at frequencies below the gap. Hence, this criterion of adiabaticity needed to be revisited. This is of special interest in connection with recent experimental observations of the Berry phase and the effect of the environment in superconducting circuits~\cite{WallraffBP,PekolaBP}.

Furthermore, the concept of a geometric phase needs to be defined (extended) for an open system, where the wave function and its evolution phases are ill-defined. While various formal generalizations have been discussed, here we take an operational approach and analyze physically relevant, directly observable quantities. Specifically, for an adiabatically manipulated isolated quantum system the Berry phase is a phase of a matrix element of the evolution operator for the wave function. For a system, weakly coupled to noise, we analyze the corresponding evolution operator of its density matrix, which determines observable quantities. We find that this operator contains phase factors, which reduce to the standard Berry phase for isolated systems.

Analysis of this kind allows one to define geometric phases for open systems, and they can be compared to their values in the limit of the vanishing coupling to the environment. An important question is whether and how are they modified by the noise.

We analyze this problem, using Bloch-type master equation for the evolution of the density matrix. Let us comment on some earlier attempts to analyze this evolution operator (cf.~the discussion in Ref.~\cite{Whitney:spin-half}). In various contexts, Bloch-type master equations have been used. However, if the standard master equation is used in the case of an arbitrary (slow) time variations of the Hamiltonian, one finds the same value of the geometric phases as without the noise. We stress, however, that the dissipative terms in the standard Bloch equations are derived for a static Hamiltonian, and they should be modified (rederived), if the Hamiltonian varies in time, even slowly. This analysis shows, that the noise-induced terms ($T_1$- and $T_2$-terms as well as the ``Lamb shift'' for a spin system) are modified by the variations of the Hamiltonian which leads to modification of the geometric phases.

In Ref.~\cite{Whitney:pre} this study has been performed for a spin-1/2 system in a magnetic field (any two-level system reduces to a spin-1/2), which varies along a cone, with the direction of the noisy contribution to the field along the axis of the cone. It was found that the Berry phase acquires a noise-induced contribution. In Ref.~\cite{Whitney:spin-half} the analysis was performed for arbitrary loops, traversed by the tip of the magnetic field, and it was found that the modification of the Berry phase is of geometric origin (similar to the Berry phase itself) and has a quadrupolar symmetry. Moreover, it was observed that this modification of the Berry phase is complex, which implies a geometric contribution to dephasing.

Here we describe another possibility to generate an environment-induced geometric phase. The analysis in Refs.~\cite{Whitney:pre,Whitney:spin-half} was done under the assumption that the spin-bath coupling is fixed. However, typically adiabatic manipulations of the Hamiltonian are performed by changing control parameters of the quantum system, and this can easily influence the coupling to the environment or the properties of the bath (cf. the effect of flux noise in Ref.~\cite{FalciBP}). The strength and the matrix structure of the system-bath coupling may also be modified deliberately. Under these circumstances new contributions to the Berry phase may arise, and the analysis of earlier work cannot be applied directly. Here we analyze the geometric phases in situations, when not only the Hamiltonian but also (or only) the coupling to the environment is varied adiabatically, and find the corresponding contributions to the geometric phase.

{\it Hamiltonian.}
A two-level system is equivalent to a spin-1/2, and we use the spin notations to describe its dynamics. The Hamiltonian of a spin coupled to a fluctuating field can be written as
\begin{eqnarray}\label{eq:Ham}
 {\cal H}=-\frac{1}{2}\bB\bS
 -\frac{1}{2}{\hat X}\bn\bS+{\cal H}_{\rm env},
\end{eqnarray}
where the field $\bB$ corresponds to the controlled part of the Hamiltonian. The second term describes the influence of the environment, the fluctuating field. To demonstrate the effect and following earlier work we consider uni-directional fluctuations, but allow the direction to vary slowly in time. Thus ${\hat X}$ is the bath operator and the slow-varying $\bn(t)$ indicates the direction and the strength of the fluctuations. The last term in Eq.~(\ref{eq:Ham}) describes the dynamics of the environment and, in particular, determines the correlation functions of ${\hat X}$. We assume that the field ${\hat X}$ has zero average (in the absence of coupling to the spin).

Variations of $\bn$ may result from the dynamics of the environment or changes in the spin-environment coupling. It may be induced deliberately or be a side effect of the adiabatic variations of the spin Hamiltonian $\bB$. To illustrate the effect, here we analyze the slow variations of the direction and the strength of a unidirectional noise with fixed correlations; this analysis can be easily generalized to include more general fluctuating fields, in particular, with a varying power spectrum\footnote{The results (\ref{Atwo}), (\ref{G}) below apply also in this case, with the substitution
$S(\Omega) \to S(\Omega,t) \equiv \int d\tau e^{i\Omega\tau} \langle \half [\hat X(t+\frac{\tau}{2}), \hat X(t-\frac{\tau}{2})]_+ \rangle$.}.

We analyze the phase accumulated by the system (and the dephasing) between times $0$ and $t_P$. Detection of this phase may involve preparation of the initial state (e.g., a superposition of $\ket{\uparrow}$ and $\ket{\downarrow}$) and the final direct or indirect measurement. We do not specify details of these events (see below). We further assume the following conditions for the time scales involved: $\tau_{\rm c}, B^{-1} \ll t_P, T_2$, where $\tau_{\rm c}$ is the noise correlation time and $T_2$ is the dephasing time scale. This implies, in particular, that the noise is weak and short-correlated~\cite{Makhlin:ext}, and that on the time scale $t_P$ of the evolution the noise correlations are local. Note also that the coherence decays on the time scale $T_2$, and at longer times $t_P\gg T_2$ the phase information is exponentially suppressed. $\bB$ and $\bn$ vary at typical frequencies $\sim\omega=2\pi/t_P$, and the adiabaticity parameter is $\omega/B$.

{\it Reference frame.}
To find the effect of the time variations of the Hamiltonian $\bB$ and the noise $\bn$, it is convenient to make a time-dependent transformation of the wave functions or, in the spin language, a transformation to a rotating reference frame (RF), in which the direction of the {\bf B}-field is stationary~\cite{Berry,Whitney:spin-half}. In that frame we use the standard procedure to write the master equation for the density matrix and thus find the evolution operator.

Below we find that the evolution of the off-diagonal entry of the density matrix in the RF, $\rho_{\uparrow\downarrow}$, is described by the expression
\begin{eqnarray}\label{Eq:BPdef}
\rho_{\uparrow\downarrow}(t)=\rho_{\uparrow\downarrow}(0) e^{i\int_0^{t}(B+i\Gamma)dt+i(\Phi^0+\delta\Phi)} \,.
\end{eqnarray}
Here the real part of the phase gives the angle of the spin precession about $\bB$, while its imaginary part (due to $\Gamma$ and $\delta\Phi$) describes the decay of the transverse spin (dephasing). In Eq.~(\ref{Eq:BPdef}) the first term in the exponent gives the dynamical phase and dephasing; this term scales with the total time $t_P$. The second term is the geometric phase, $\Phi_0+\delta\Phi$, insensitive to time reparameterization: the bare Berry phase $\Phi_0$ depends only on the geometry of the path $\bB(t)$, and the environment-induced complex contribution $\delta\Phi$ depends on the geometry of $\bB(t)$ and $\bn(t)$.

The analogy with the spin dynamics shows that, similar to the Berry phase for a closed system, the (total and) geometric phase at time $t_P$ for an open system is meaningful only when the direction of the field $\bB$ at $t_P$ coincides with its initial value (however, there is no restriction on $\bn(t_P)$). At the same time, the imaginary part of the phase, the decay of coherence, is well-defined also for open paths (cf.~Ref.~\cite{Whitney:spin-half}).

To choose the frame, we choose eigenstates $\ket{\uparrow_t}$, $\ket{\downarrow_t}$ of $\bB(t)\bS$ for each $t$ (we omit the subindex $t$ below). In the spin language this fixes the axes of the RF: $\hat z = \bB/B$, while $\hat x$ is such that $\ket{\uparrow_t} + \ket{\downarrow_t}$ is the $+1$-eigenstate of $\hat x\bS$. We have the freedom to choose the phases of the states $\ket{\uparrow/\downarrow}$, or the direction of the $x,y$-axes; this choice does not affect the results, we only assume that the states (the axes) vary slowly enough, at frequencies $\ll B$; we further suppose that the $x,y$-axes assume their initial values if we consider a closed loop, i.e., when the direction of $\bB(t)$ at the final time returns to its initial value.

In this rotating frame the pseudo-magnetic field is $\bB' = \bB + \bomega$, where $\bomega$ is the angular velocity of the RF. The direction of $\bB'$ is stationary to the leading order in the adiabatic parameter but differs from the $z$-axis, and it is convenient to choose it as the $z'$-axis in the RF (``$x'y'z'$-frame''). This transformation of the basis in the RF, which changes the direction of the third axis from $\bB$ to $\bB'$, is only weakly time-dependent ($\dot{\bomega}\sim\omega^2$), and thus corrections to the pseudo-magnetic field at this step are negligible.

The eigenstates of the Hamiltonian in the RF (that is, the eigenstates of $\bB'\bS=B'\sigma_{z'}$) can be easily found as
\begin{eqnarray}\label{newbasis}
\ket{\uparrow^\prime}&=&\ket{\uparrow}+i\frac{\braket{\downarrow}{\dot\uparrow}}{B}\ket{\downarrow} \,,\\
\ket{\downarrow^\prime}&=&\ket{\downarrow}-i\frac{\braket{\uparrow}{\dot\downarrow}}{B}\ket{\uparrow} \,,\nonumber
\end{eqnarray}
(notice the choice of the overall phases).

Using these eigenstates we find the level splitting in the RF:
\begin{equation}\label{eq:Bprime}
B'\approx B+i \left(\braket{\uparrow}{\dot\uparrow} - \braket{\downarrow}{\dot\downarrow}\right) \,.
\end{equation}
To proceed with the calculation in the RF, we need to find the components of the fluctuating field. We introduce the coordinates $n_z$, $n_\pm = n_x\pm i n_y$ of the $\bn$-vector in the $xyz$-frame: $\bn\bS = \half n_+\sigma_- + \half n_-\sigma_+ + n_z\sigma_z$, where $\sigma_\pm=\sigma_x\pm i\sigma_y$. One finds that its coordinates in the primed frame are slightly different, and to the leading order in $\omega$:
\begin{eqnarray}
n'_+-n_+ &=& - \frac{2i n_z}{B}\braket{\downarrow}{\dot\uparrow} \,,\\
n'_z-n_z &=& \frac{i}{B}
   \left(n_+\braket{\uparrow}{\dot\downarrow}+
   n_-\braket{\downarrow}{\dot\uparrow}\right)\,,
\end{eqnarray}
and $n'_-=(n'_+)^*$.

Below we find that under the specified conditions the equation of motion for the off-diagonal entry $\rho_{\uparrow\downarrow}$ of the density matrix in the primed basis is decoupled from the other matrix elements, and the evolution factor contains information about the Berry phase~\cite{multi}. This factor can be extracted from the measurements of the spin state.

{\it Evolution of the density matrix}.
To calculate $\rho_{\uparrow\downarrow}(t)$, we derive a Bloch-Redfield master equation for the evolution of the reduced density matrix of the two-level system considered (cf., e.g., Refs.~\cite{Bloch_Derivation,Redfield_Derivation,GardinerZoller,Makhlin:ext} and references therein). Let us outline the derivation: we begin from the Liouville equation for the density matrix of the combined system and environment, assuming an initially factorized density matrix, then make the transformation to the interaction representation, and expand the evolution operator to the second order in the system-bath coupling. The derivation is performed under the assumption of weak and short-correlated noise, which allows one to reduce the integro-differential equation of motion to a markovian differential equation. Finally, the secular (or rotating-wave) approximation allows us to decouple the evolution of different entries of the density matrix.

After the expansion in the perturbation ${\cal V}' = -\frac{1}{2}{\hat X}\bn'\bS'$ and averaging over the state of the environment, we find:
\begin{eqnarray}
&&(i\partial_{t}+B')\rho_{\uparrow\downarrow}(t) =
\nonumber\\
&&= -i \int^t_{-\infty}
\left\langle\left[\left[\ket{\downarrow_t}\bra{\uparrow_t},{\cal V}'(t) \right],
{\cal V}'(t_1)\right]\right\rangle dt_1.
\label{halfmaster}
\end{eqnarray}
(We can set the lower limit of integration to $-\infty$ for $t\gg\tau_{\rm c}$, noise correlation time, which is also the convergence scale of the integral in Eq.~(\ref{halfmaster})~\cite{Bloch_Derivation,Redfield_Derivation,GardinerZoller}; the $\sim\tau_{\rm c}$-intervals at the boundary contribute to the boundary phase $\delta\Phi_{\rm b}$, see below.)

Using the secular approximation ($\Gamma\ll B$), we obtain:
\begin{eqnarray}
(i\partial_t+B')\rho_{\uparrow\downarrow} = -i \rho_{\uparrow\downarrow}\int_{-\infty}^t S(t-t_1) \times\nonumber\\
\left(\half n'_-(t)n'_+(t_1)e^{-i\int_{t_1}^t B'(\tau) d\tau} + n'_z(t)n'_z(t_1)\right)dt_1
\,,\label{halfmaster2}
\end{eqnarray}
where $S(t-t_1)=\frac{1}{2}\langle{\hat X}(t){\hat X}(t_1)+{\hat X}(t_1){\hat X}(t)\rangle$ is the symmetrized correlation function of the noise.

For an isolated system, i.e., neglecting the rhs of Eq.~(\ref{halfmaster2}), one finds the phase acquired as
$\Phi_{\rm total}(t_P)=\int_0^{t_P} B' dt=\int_0^{t_P} (B+(\bomega\bB)/B)dt=\int_0^{t_P} Bdt+\Phi_{BP}^0$, which is the sum of the dynamical phase and the conventional Berry phase. $-\Phi_{BP}^0 \pmod{2\pi}$ is given by the solid angle subtended by the path $\bB(t)$.

The fluctuations on the rhs of Eq.~(\ref{halfmaster2}) give rise to the dephasing of the element $\rho_{\uparrow\downarrow}$ and the noise-induced contribution to the phase. If we neglect all order-$\dot\bB, \dot\bn$ effects, this expression produces terms $\propto t$ in the total phase: setting $n_-^\prime(t)=n_-(t)=n_-(t_1)$, $n_+^\prime(t)=n_+(t)$, and $B^\prime=B$, we find:
\begin{eqnarray}\label{Gtwo}
\Gamma=i\left(\frac{|n_+|^2}{2} \int\frac{d\Omega}{2\pi}\frac{S(\Omega)}{\Omega-B+i0}
+n_z^2\int\frac{d\Omega}{2\pi}\frac{S(\Omega)}{\Omega+i0}\right).
\end{eqnarray}
Here $S(\Omega)$ is the Fourier transform of $S(t)$.
Note that $\Gamma$ is complex, its real part gives the decoherence rate, while its imaginary part determines the modification of the level splitting (``Lamb shift''), cf.~Ref.~\cite{Makhlin:ext}.

To find the noise-induced modification of the Berry phase, we expand the rhs of Eq.~(\ref{halfmaster2}) to the first order in $\dot{\bB}$ and $\dot\bn$. Integrating the evolution over time, we find a geometric contribution:
\begin{eqnarray}
&&\delta\Phi=\nonumber\\
&&\int\left( i\frac{S(0)}{B}
-\int\frac{d\Omega}{4\pi}\frac{S(\Omega)(3B-2\Omega)}{B(\Omega-B+i0)^2}
\right)
\frac{\bn\bB}{B}\frac{\bn(\bB\times d\bB)}{B^2}
\nonumber\\
&&-\frac{1}{2}\int\left(\int\frac{d\Omega}{2\pi}\frac{S(\Omega)}{(\Omega-B+i0)^2}\right)
\frac{\bB(\bn\times d\bn)}{B}
\nonumber\\
&&+\int dG
+\delta\Phi_{\rm b}
\,.\label{Atwo}
\end{eqnarray}
Here the quantity $G$ is given by
\begin{eqnarray}\label{G}
&&G(\bn,B)= G(|n_+|,n_z,B)=\\
&&i\frac{|n_+|^2}{4}\int\frac{d\Omega}{2\pi}\frac{S(\Omega)}{(\Omega-B+i0)^2}
+i\frac{n_z^2}{2}\int\frac{d\Omega}{2\pi}\frac{S(\Omega)}{(\Omega+i0)^2}
\,.\nonumber
\end{eqnarray}
The last term $\delta\Phi_{\rm b}$ is the boundary contribution and will be discussed below.
Eq.~(\ref{Atwo}) is our main result.

Let us discuss various terms in Eq.~(\ref{Atwo}).
The first term gives the contribution due to variation of the Hamiltonian $\bB$ at fixed $\bn$, which was found in Ref.~\cite{Whitney:spin-half}. Here we present it in a form, independent of a choice of coordinates, which may be convenient for applications. We see from this expression that this contribution arises, when the vector of the pseudo-magnetic field $\bB$ rotates around $\bn$.

{\it Geometric phase generated by the rotation of noise.}
The next term of Eq.~(\ref{Atwo}), $\int\ldots d\bn$, describes the correction to the geometric phase due to the variations in the system-environment coupling. It appears when the direction $\bn$ of the fluctuating field rotates around the bare field $\bB$. We refer to this term as the term due to rotation of the noise.

\begin{figure}
  \centerline{{\sf a}\includegraphics[width=1in]{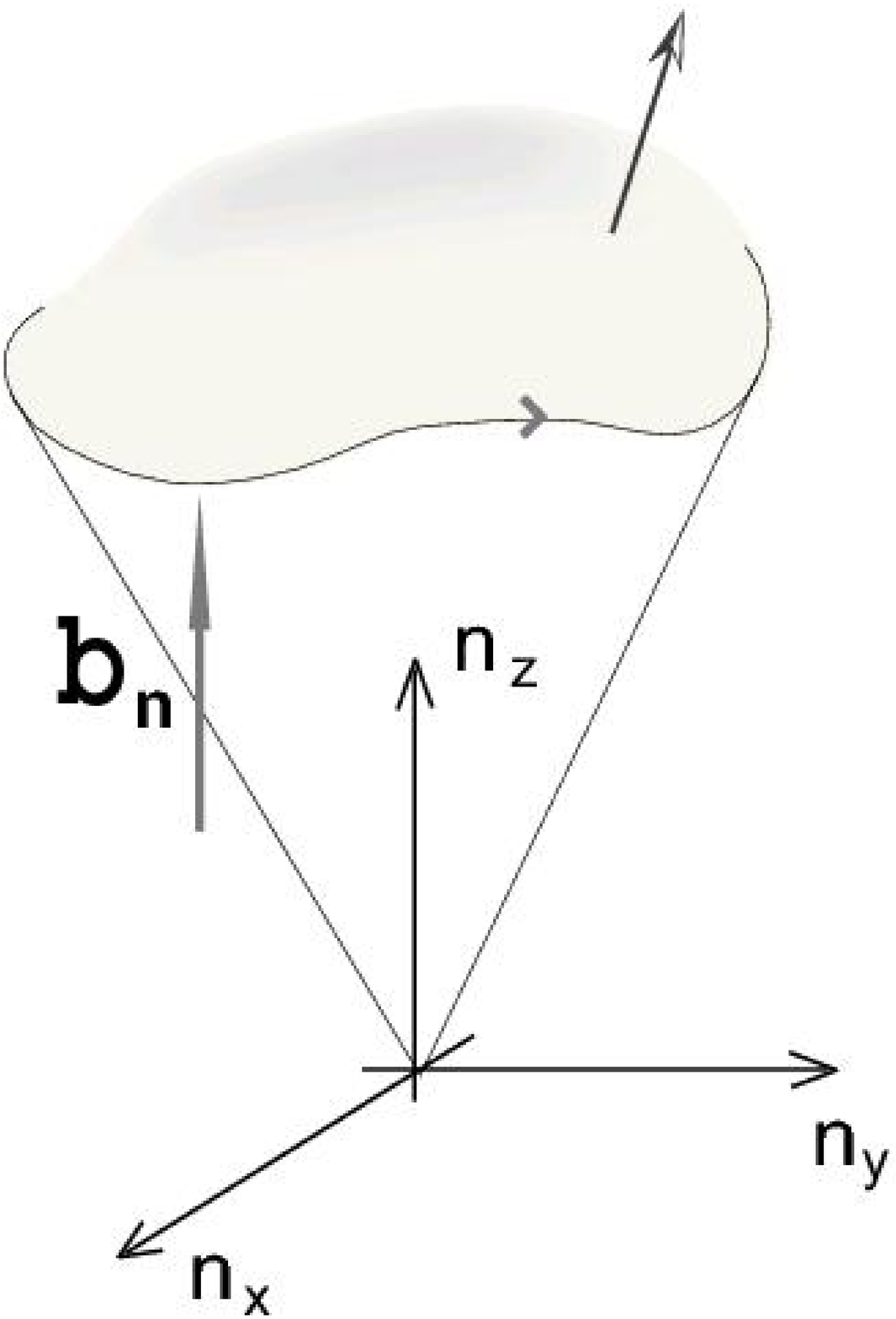}\qquad{\sf b}\includegraphics[width=1.7in]{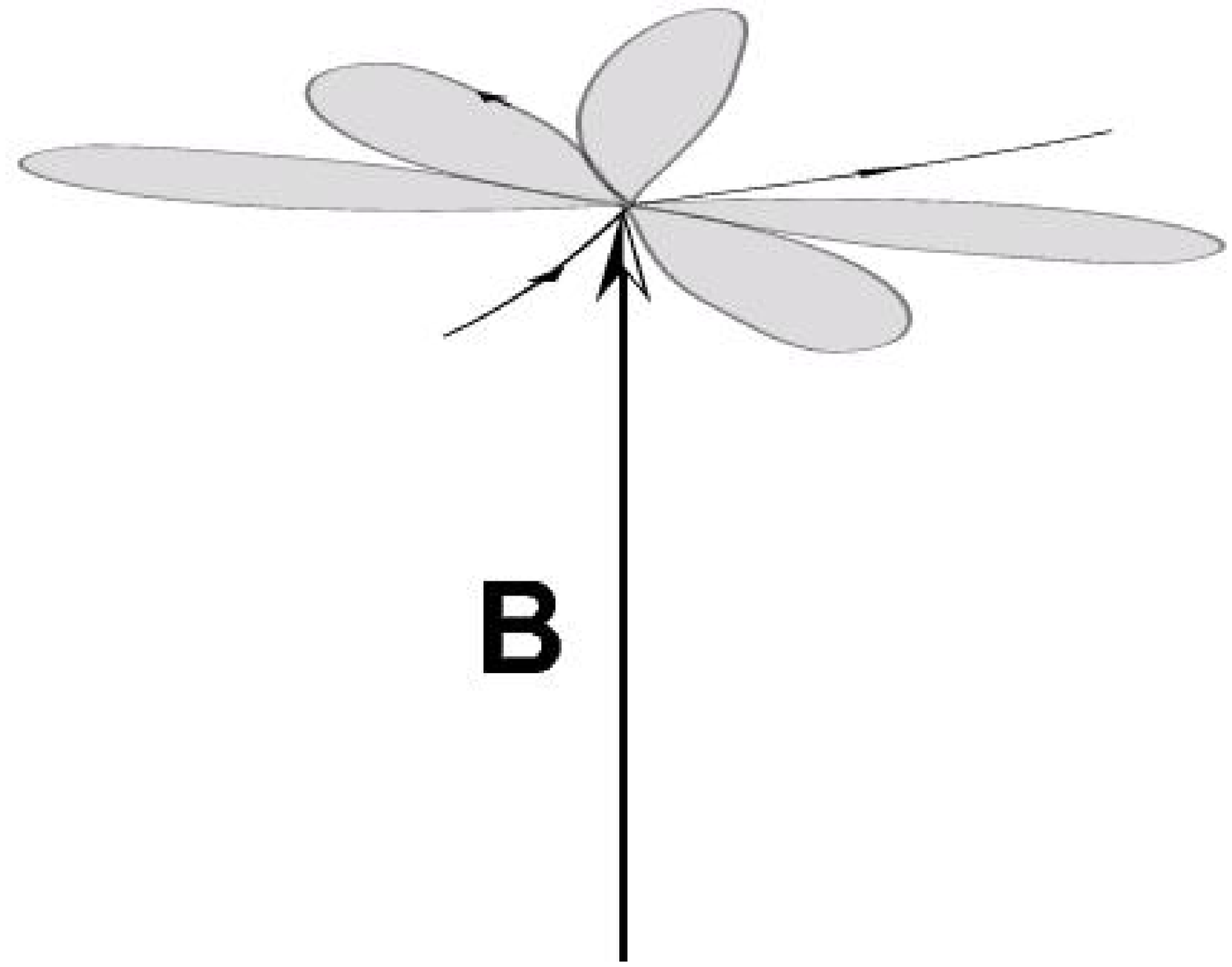}}
  \caption{\label{Fig:surfnoiserot}FIG.\ref{Fig:surfnoiserot}.
  a) The rotation of noise about a constant magnetic field $\bB$
  generates a geometric phase given by the flux of
  a uniform field ${\bf b_n}$ through the loop $\bn(t)$.
  \label{Fig:swept}
  b) The area swept by the vector $\bB+\bn(t)X(t)$ in the plane, perpendicular to $\bB$}
\end{figure}

Let us illustrate the origin of this term by considering the simplest example, in which this contribution arises: the bare Hamiltonian retains its initial value, $\bB={\rm const}$, whereas the direction $\bn$ of noise varies in the plane, transverse to $\bB$, with angular velocity $\dot\varphi$ (Fig.\ref{Fig:swept}b). Then according to Eq.~(\ref{Atwo}) the first term vanishes. We postpone the discussion of the last line in this equation and consider the contribution due to rotation of noise. The considerations are simplified by going to the frame, rotating together with $\bn$ around $\bB$. In that frame $\bn$ is constant; the field $\bB$ is also constant but it acquires an extra contribution, $B \to B+\dot\varphi$. This implies that the rate $\Gamma$ (\ref{Gtwo}) is modified compared to its value in the laboratory frame, and hence the total phase acquires a contribution $i\int\delta\Gamma dt=i\int(\partial_B\Gamma) \dot\varphi dt$:
\begin{eqnarray}\label{noiserot}
\Phi_{\rm geom}^{\rm noise}=\oint i\frac{\partial\Gamma}{\partial B}d\varphi \,.
\end{eqnarray}

The phase $\Phi_{\rm geom}^{\rm noise}$ has the following geometric interpretation: for a closed loop $\bn(t)$ and $\bB={\rm const}$ this phase is given by the flux of the uniform field (see Fig.\ref{Fig:surfnoiserot}a)
\begin{eqnarray}\label{nrfield}
{\bf b_\bn}=-{\bf e_z}\int\frac{d\Omega}{2\pi}\frac{S(\Omega)}{(\Omega-B+i0)^2}
\end{eqnarray}
through the surface spanned by the loop $\bn(t)$.
This result follows directly from the Eq.~(\ref{Atwo}).

These results follow immediately from the well-known expression of the Berry phase via the solid angle, if the noise is slow (typical $\Omega\ll B$). Indeed, during the rotation, transverse adiabatic fluctuations of the total magnetic field $\bB+X\bn$ sweep the area $\half|\bn_\bot|^2\langle X^2\rangle\dot\varphi =\half |n_+|^2\langle X^2\rangle\dot\varphi$ per unit time (see Fig.\ref{Fig:swept}b). Thus the geometric phase is:
\begin{eqnarray}
\Phi^{\rm rot}_{\rm slow\ noise}=-\frac{1}{2}|n_+|^2\langle X^2\rangle\int\frac{d\varphi}{B^2} \,,
\end{eqnarray}
in agreement with Eq.~(\ref{Atwo}).

Thus, we find that the variation of the system-bath coupling may generate a geometric phase, that is, a phase which depends only on the geometry of the path $\bn(t)$. This phase has both real and imaginary parts, that is, it describes a geometric contribution to dephasing as well.

One can also understand the contributions in Eq.~(\ref{Atwo}) in the following way: if neither $\bB$ nor $\bn$ rotate about each other, i.e., when the $(\bB,\bn)$-plane retains its direction, the contribution to the geometric phase (due to variations of $\bB$ and $\bn$ within the plane, that is, due to changes of their magnitudes and the relative angle) can only be of potential nature~\cite{Whitney:spin-half} and give the term $\int dG$ in Eq.~(\ref{Atwo}). The generic case can always be reduced to it by going to a proper rotating frame; for instance, one can begin with the RF used in the calculations above and add an additional rotation about $\bB'$ to keep the phase of $n_+$ constant.  This additional rotation, with the angular velocity $\dot\varphi$ (here $n_+=|n_+|e^{i\varphi}$), modifies the field as $B'\to B'+\dot\varphi$ and thus the dephasing rate in Eq.~(\ref{Gtwo}), producing a contribution to the phase $\delta \Phi\leftarrow i\int d\varphi \partial_B\Gamma$, which coincides with the second term of Eq.~(\ref{Atwo}) (cf.~also (\ref{noiserot})).

{\it The boundary phase.}
The last term in Eq.~(\ref{Atwo}) is a boundary contribution and is accumulated in the vicinity of the initial and the final points, $t=0$ and $t=t_P$; it cannot be presented as a line integral. Let us comment on the origin and the value of this term. In the derivation of the master equation~(\ref{halfmaster}) we assumed a factorized density matrix of the system~+~environment at an earlier time $t_0=-\infty$, the latter served as the lower limit of integration in Eq.~(\ref{halfmaster}). In fact, the exact initial conditions for the density matrix and, in particular, the exact value of $t_0$ do not matter except for times $t$ very close to $t_0$, in an interval $\sim\tau_{\rm c}$. However, the behavior of the density matrix and the phase accumulated within this short time interval are sensitive to the initial conditions; this contribution to the phase depends on the details of the preparation at the beginning of the adiabatic manipulations and does not scale with the total time $t_P$ of the Berry-phase experiment, hence this boundary effect should be formally ascribed to the geometric phase. The same considerations apply to the short final interval, when the (quantum) measurement of the final state is performed, the exact value of the phase being sensitive to the details of the read-out procedure. These two contributions form the boundary term $\delta\Phi_{\rm b}$. For an abrupt measurement at $t=t_P$ the final boundary term vanishes; the initial boundary contribution also vanishes for the slow or preliminary preparation, which corresponds to $t_0=-\infty$; in contrast, for an abrupt preparation with $t_0=0$ one finds from Eq.~(\ref{halfmaster}) that $\delta\Phi_{\rm b} = 2 G(\bn(0),\bB(0))$.

Notice that the boundary phase masks the other contributions in Eq.~(\ref{halfmaster}). For a closed path, as emphasized in Ref.~\cite{Whitney:spin-half}, the contribution of the first terms can be enhanced relative to the boundary term by traversing the path several times, $N$: their contribution scales $\propto N$, whereas $\delta\Phi_{\rm b}$ retains its value. However, the term $\int dG$ vanishes for a closed loop and cannot be enhanced by this method. In other words, to calculate and compare to experiment the last line in Eq.~(\ref{halfmaster}) one needs to take into account the details of the preparation and read-out.

{\it In conclusion}, we found that slow variation of the properties of the noise or of the system-environment coupling may result in a geometric phase. We found this contribution for a two-level system in a fluctuating field with a slowly varying direction. We thank A.~Shnirman for discussions. This work was partially supported by the projects INTAS 05-1000008-7923, MD-4092.2007.2, and the Dynasty foundation.

\end{document}